**Changes in porosity, permeability and surface area during rock dissolution: effects of mineralogical heterogeneity**

Ting Min [a], Yimin Gao [a], Li Chen [b, c, *], Qinjun Kang [c], Wen-Quan Tao [b]

a: State Key Laboratory for Mechanical Behavior of Materials, Xi'an Jiaotong University, Xi'an 710049, PR China

b: Key Laboratory of Thermo-Fluid Science and Engineering of MOE, School of Energy and Power Engineering, Xi'an Jiaotong University, Xi'an, Shaanxi 710049, China c: Computational Earth Science, Earth and Environmental Sciences Division, Los Alamos National Laboratory, Los Alamos, New Mexico 87545, USA

* Corresponding author: lichenmt@lanl.gov, lichennht08@mail.xjtu.edu.cn

**Abstract:**

Effects of heterogeneity of mineral distribution and reaction rate on the rock dissolution process are investigated using a pore-scale reactive transport model based on the lattice Boltzmann method. Coupled fluid flow, species transport, chemical reaction and solid structure alternation due to dissolution are simulated. Effects of mineral distributions and chemical heterogeneity on the dissolution behaviors and evolutions of hydrologic properties are studied under different reactive transport conditions. Simulation results show that the coupling between advection, diffusion and reaction as well as the mineralogical heterogeneity leads to complex reactive transport behaviors and complicated temporal evolutions of hydrologic properties including porosity, permeability and reactive surface. Diverse relationships between surface area and volume are predicted, which cannot be described by simple models such as the spherical-grain model. Porosity-permeability relationships also differ under different mineral distributions and reactive transport conditions. Simulation results indicate that it is extremely challenging to propose general relationships for hydrologic properties for dissolution of rocks with mineralogical heterogeneity, due to the complicated interactions between reactive transport and mineralogical heterogeneity.

## 1. Introduction

Reactive mass transport in heterogeneous porous media with solid phase dissolution is ubiquitous in geological formations, scientific processes and industrial application [1-4]. Typical examples include karst formation [5], self-assembled patterns [6], spread of contaminants in fluid-saturated soils [7], nuclear waste remediation [8], geologic sequestration of carbon dioxide [9-12] and acid injection for enhanced petroleum recovery [13]. The heterogeneity of porous media in geological formations is embodied by the heterogeneous porous structures [14-22] as well as the mineralogical heterogeneity because of multiple components [23-27]. In such heterogeneous porous media, the observed reactive transport processes do not always behave according to the transport laws established for the homogeneous ones [14-27]. For continuum-scale reactive transport modeling applied at large scales, the physicochemical heterogeneities are necessarily ignored at scales smaller than the size of the model discretization [4]. However, while under same scenarios the homogeneous assumption is reasonable, the pore-scale heterogeneities can result in significant "scaling effect" because of the spatial variations of concentrations and reaction rates, leading to the breakdown of the homogeneous assumption. Such "scaling effect" is one of the potential causes of the order-of-magnitude differences between lab measured reaction rates and that obtained from the field measurements [1, 4]. Therefore, it is of significant importance to understand the effects of pore-scale heterogeneities on the reactive transport processes.

Theoretically, full resolutions of the spatial medium heterogeneity as well as the detailed mineral distributions down to the pore scale, with all the physicochemical processes considered, can help to understand the distinct reactive transport phenomena, to establish the reactive transport laws, and to reveal coupled mechanisms in heterogeneous porous media. The transport of a reactive fluid through a porous medium with dissolution is a very complex process encompassing multiple physicochemical sub-processes including fluid flow, species transport, chemical reactions, and alternations of solid and porous structures [14-27]. These sub-processes occur simultaneously and are closely coupled with each other. With the improvement of the computational resources, pore-scale modeling and simulations have been developed as a powerful tool for studying such reactive transport processes. A desirable pore-scale reactive transport model must be able to address the multiple physicochemical sub-processes. Various numerical methods have been adopted to model the fluid flow and species transport with chemical reactions, such as direct numerical simulations [14], pore-network modeling [27], smooth particle hydrodynamics [15], and the lattice Boltzmann method [3, 8, 19, 21, 22, 24, 28, 29]. Further, solid structure alternations resulting from dissolution have also been addressed by different interface capturing/tracking models such as the phase field method [30], the cellular automaton method [31], the volume of fluid model and the level set method [32, 33]. There are two major objectives for pore-scale reactive transport modeling. One is to understand the underlying reactive transport phenomena and to reveal the coupled mechanisms between different processes. Pore-scale simulations have revealed a complex coupling between convection, diffusion and dissolution reaction under the effects of structural heterogeneity [3, 14, 18-20, 22]. Different dissolution patterns have been found including uniform dissolution, face dissolution and wormhole dissolution under different reactive transport conditions. The other purpose is to determine the hydrologic properties of the porous medium such as porosity, permeability and reactive surface area, which are required in continuum-scale models [34]. Taking porosity-permeability relationship as an example, under the influence of heterogeneity, the complex interactions between dissolution and reactive transport will generate different dissolution patterns, and thus lead to quite different porosity-permeability curves [9, 22].

Although some progresses have been made in the understanding of the coupling of dissolution reaction and fluid flow in porous rocks, there are still some fundamental problems remaining to be solved. First, while there have been some pore-scale studies regarding the evolutions of porosity and permeability during the dissolution [3, 15, 19, 22, 33, 35], the work about the evolutions of the reactive surface area is scarce. In continuum-scale reactive transport models, the reactive surface area, which is not easy to measure during experiments, is an important prerequisite for calculating the dissolution rate [36, 37]. Second, subsurface porous media consist of multiple mineral components such as calcite, clay, quartz, dolomite and pyrite [24-26, 37]. Such mineralogical heterogeneity also plays an important role on the dissolution process [24-26], which, however, was usually ignored in the open literature about pore-scale simulations of reactive transport with dissolution. In our previous study [24], we found that the undissolved mineral causes heterogeneous local dissolution behaviors, and leads to porosity-permeability curves that significantly differ from that in a mono-mineral system [24]. Wormhole formation also could be suppressed by the undissolved mineral [24]. Further studies are required to investigate the effects of spatial and chemical heterogeneities in multiple mineral porous systems.

In the present study, reaction transport with dissolution in binary-mineral rocks is simulated. Effects of mineral distributions and orientations on dissolution processes are explored. Time evolutions of important hydrologic parameters including porosity, permeability and reactive surface area are monitored and discussed. The relationship between reactive surface area and solid volume, and that between porosity and permeability are presented and discussed. A LB based pore-scale reactive transport model is adopted for the present study, which has been well developed for single-phase reactive transport with dissolution-precipitation [28, 38, 39] and has been extended for multiphase reactive transport very recently [8, 12]. As a first step study to reveal the complexity of effects of mineral distributions and reactive transport conditions on dissolution process as well as evolutions of hydrologic properties, an idealized structure with relatively simple mineral distributions are studied. As will be shown in the results, even for such a simple idealized structure, the evolutions of the hydrologic properties are very complicated and it is challenging to obtain a general relationship. The remainder of present paper is organized as following. In section 2, physicochemical model is established and the LB reactive transport pore-scale model is introduced. The results about dissolution in the idealized structure as well as that in a system with relatively complex mineral distribution are presented and discussed in Section 3. Finally, some conclusions are drawn in section 4.

## 2. Physicochemical models and Numerical methods

2.1 Physicochemical models

It is ubiquitous that porous rock systems consist of multiple minerals, which have distinct chemical properties and spatial distributions [25, 37]. During reactive mass transport these minerals existing in porous rock may or may not be dissolved by chemical fluid surrounding them [25]. To emphasis the effects of the chemical heterogeneity of rocks, it is assumed that there are two minerals in rocks: one soluble mineral, $\alpha$, and the other insoluble mineral, $\beta$ [24]. Without loss of generality, a simplified dissolution reaction of $\alpha$ is considered in the present study [3]

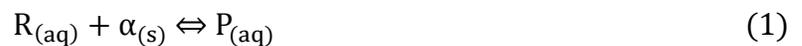

$$R_{(aq)} + \alpha_{(s)} \Leftrightarrow P_{(aq)} \tag{1}$$

with $R_{(aq)}$ is the reactant in aqueous phase, $\alpha_{(s)}$ the reactant in solid form and $P_{(aq)}$ the product in aqueous phase. An example of this kind of system is the limestone in acid fluid, which consists of calcite and clay, and calcite can react with acid solution and clay cannot [25]. It is assumed that the dissolution reaction is the first-order chemical reaction, and the dissolution rate can be calculated by [3]

$$r = k_r(C_{R_{(aq)}} - C_{P_{(aq)}}/k_{eq}) \tag{2}$$

with $r$ [mol m$^{-2}$ s$^{-1}$] the dissolution rate, $k_r$ [m s$^{-1}$] the dissolution reaction rate constant, $C_R$ the reactant concentration, $C_P$ the product concentration and $k_{eq}$ [mol m$^{-3}$] the reaction equilibrium constant.

In our simulations, it is assumed that the solute transport has negligible influence on the fluid flow, which is reasonable when the solute concentration is low enough. The incompressible fluid flow is driven by a constant pressure difference and complies the following governing equation

$$\frac{\partial \rho}{\partial t} + \nabla \cdot (\rho \mathbf{u}) = 0 \tag{3a}$$

$$\frac{\partial \mathbf{u}}{\partial t} + (\mathbf{u} \cdot \nabla)\mathbf{u} = -\frac{1}{\rho}\nabla p + \nabla \cdot (\nabla \nu \mathbf{u}) \tag{3b}$$

with $\rho$ fluid density, $t$ time, $\mathbf{u}$ velocity and $\nu$ kinematic viscosity. The solute transport includes advection and diffusion and can be described as

$$\frac{\partial C_k}{\partial t} + (\mathbf{u} \cdot \nabla)C_k = D_k \Delta C_k \tag{4}$$

with $C_k$ the concentration of species $k$, and $D_k$ the diffusivity of species $k$. At the interface between fluid and α mineral, reactant R$_{(aq)}$ is consumed while product P$_{(aq)}$ is generated through the dissolution reaction Eq. (1) and the boundary condition are as follows

$$D_{R_{(aq)}}\frac{\partial C_{R_{(aq)}}}{\partial n} = k_r\left(C_{R_{(aq)}} - \frac{C_{P_{(aq)}}}{k_{eq}}\right) \tag{5a}$$

$$D_{P_{(aq)}}\frac{\partial C_{P_{(aq)}}}{\partial n} = -k_r\left(C_{R_{(aq)}} - \frac{C_{P_{(aq)}}}{k_{eq}}\right) \tag{5b}$$

with $n$ the direction normal to the reactive interface and pointing to the fluid. When the dissolution reaction proceeds, the volume of mineral α at the interface nodes is decreased according to

$$\frac{\partial V_{\alpha_{(s)}}}{\partial t} = -S_n M_\alpha k_r\left(C_{R_{(aq)}} - \frac{C_{P_{(aq)}}}{k_{eq}}\right) \tag{6}$$

with $V_\alpha$ the volume of mineral α at an interface node, $S_n$ the reactive surface area of the node and $M_\alpha$ the molar volume of mineral α.

2.2 LB model for Fluid flow

The LBM is a powerful tool to simulate the reactive transport process involving interfacial dynamics and complicated boundary conditions, like multiphase or multicomponent flow in porous media due to its kinetic nature [40-42]. In this literature, an incompressible LB model proposed by Guo et al. [43] is adopted for simulating fluid flow, in which the compressible effects of traditional LB model are avoided, and thus, it is widely applied to the incompressible Newton fluid. This evolution equation of the incompressible LB model can be described as[43]

$$f_i(\mathbf{x} + c\mathbf{e}_i \Delta t, t + \Delta t) = f_i(\mathbf{x}, t) - \frac{f_i(\mathbf{x}, t) - f_i^{eq}(\mathbf{x}, t)}{\tau_f}, i = 0,1,\cdots,8 \quad (7)$$

with $f_i(\mathbf{x},t)$ the density distribution function at lattice site $x$ and time $t$, $\Delta t$ time increment, $c$ lattice velocity equaling to $\Delta x/\Delta t$, $f_i^{eq}(\mathbf{x},t)$ the equilibrium distribution function, $\tau_f$ relaxation time and $\mathbf{e}_i$ discretized velocities. For the D2Q9 scheme (two dimensional nine-velocity) employed here, the discretized velocities $e_i$ are defined as

$$\mathbf{e}_i = \begin{cases} 0, & i = 0 \\ \left(\cos\frac{(i-1)\pi}{2}, \sin\frac{(i-1)\pi}{2}\right), & i = 1,2,3,4 \\ \left(\sqrt{2}\cos(\frac{(i-5)\pi}{2} + \frac{\pi}{4}), \sqrt{2}\sin(\frac{(i-5)\pi}{2} + \frac{\pi}{4})\right), & i = 5,6,7,8 \end{cases} \quad (8)$$

The equilibrium distribution function in the incompressible LB model for D2Q9 scheme is the predominant improvement comparing to the traditional model

$$f_i^{eq}(\mathbf{x}, t) = \begin{cases} -4\sigma\frac{P(\mathbf{x},t)}{c^2} + s_i(\mathbf{u}) & i = 0 \\ \lambda\frac{P(\mathbf{x},t)}{c^2} + s_i(\mathbf{u}) & i = 1,2,3,4 \\ \gamma\frac{P(\mathbf{x},t)}{c^2} + s_i(\mathbf{u}) & i = 5,6,7,8 \end{cases} \quad (9)$$

Here, $\sigma$, $\lambda$, $\gamma$ are parameters satisfying with two conditions: $\lambda + \gamma = \sigma$ and $\lambda + 2\gamma = 1/2$. $P(x,t)$ is pressure and $s_i(\mathbf{u})$ can be described as,

$$s_i = w_i \left( \frac{3c\mathbf{e}_i \cdot \mathbf{u}}{c^2} + \frac{9(c\mathbf{e}_i \cdot \mathbf{u})^2}{2c^2} - \frac{3\mathbf{u} \cdot \mathbf{u}}{2c^2} \right), i = 0,1,\cdots,8 \tag{10}$$

with $w_i$ the weight coefficient, which equals to 4/9 for $i = 0$, 1/9 for $i = 1, 2, 3, 4$, and 1/36 for $i = 5, 6, 7, 8$. The density and velocity can be calculated by the first and second moments of density distribution function

$$\rho(\mathbf{x},t) = \sum_i f_i(\mathbf{x},t) \tag{11a}$$

$$\rho(\mathbf{x},t)\mathbf{u}(\mathbf{x},t) = \sum_i c\mathbf{e}_i f_i(\mathbf{x},t) \tag{11b}$$

The kinetic viscosity $\upsilon$ can be determined by relaxation time $\tau_f$,

$$\upsilon = \frac{c^2}{3}(\tau_f - 0.5)\Delta t \tag{12}$$

2.3 LB model for mass transfer

The mass transfer can be simulated by D2Q5 scheme, which has comparable accuracy with much less CPU time comparing with D2Q9 [6, 24, 44]. The evolution equation of distribution function is as follows

$$g_{i,k}(\mathbf{x} + c\mathbf{e}_i \Delta t, t + \Delta t) = g_{i,k}(\mathbf{x},t) - \frac{g_{i,k}(\mathbf{x},t) - g_{i,k}^{eq}(\mathbf{x},t)}{\tau_D} \tag{13}$$

with $g_{i,k}(\mathbf{x},t)$ the distribution function of the $k$-th species at lattice site $x$ and time $t$. $g_{i,k}^{eq}$ is the equilibrium distribution function for $k$-th species

$$g_{i,k}^{eq}(\mathbf{x},t) = C_k(\mathbf{x},t)\left(J_{i,k} + \frac{1}{2}c\mathbf{e}_i \cdot \mathbf{u}\right), i = 0,1,2,3,4 \tag{14}$$

with $J_{i,k}$ given by

$$J_{i,k} = \begin{cases} J_{0,k}, & i = 0 \\ \frac{(1-J_{0,k})}{4}, & i = 1,2,3,4 \end{cases} \tag{15}$$

The diffusivity $D$ varies with $J_{i,k}$ and relaxation time $\tau_D$ by

$$D_k = \frac{1}{2}(1 - J_{0,k})(\tau_D - 0.5) \tag{16}$$

Thus, various diffusivities can be obtained by changing $J_{0,k}$ or $\tau_D$. The concentration can be got by summing the distribution function

$$C_k(\mathbf{x}, t) = \sum_i g_{i,k}(\mathbf{x}, t) \tag{17}$$

2.4 LB boundary conditions for surface reaction

The reactive boundary conditions described in Eq. (5) need to be transformed into expression in the LB model. Several LB boundary conditions have been developed for surface reaction [8, 44-47]. The one proposed by Kang et al. [44] is adopted. First, the flux at the solid-liquid interface is as follows

$$\sum g_i e_i = cu - D\nabla c \tag{18}$$

In our simulation, the solid blocks are static, which means the velocity of interface node is zero, thus, the first term on the right in the above equation is zero, too. Combining the Eq. (5), the following formula is obtained

$$\sum g_i e_i = -D\nabla c_R = -k_r \left( C_{R_{(aq)}} - \frac{C_{P_{(aq)}}}{k_{eq}} \right) \tag{19a}$$

for reactant R, and

$$\sum g_i e_i = -D\nabla c_p = k_r \left( C_{R_{(aq)}} - \frac{C_{P_{(aq)}}}{k_{eq}} \right) \tag{19b}$$

for product P. For static wall, Kang [44] assumed that the non-equilibrium portions of the distribution functions in opposite directions have the same value with opposite sign. For example, for a horizontal wall

$$g_2 + g_4 = g_2^{eq} + g_4^{eq} = \frac{1 - J_{0,k}}{2} C \tag{20}$$

Combining Eqs. (19) and (20), the unknown distribution functions at the interface nodes can be solved.

2.5 Update of the solid structures

Mineral α dissolves due to the dissolution reaction, and its volume changes according to Eq. (6). Eq. (6) is updated at each time step as follows [28]

$$V_\alpha(t + \Delta t) = V_\alpha(t) - S_n M_\alpha k_r \left( C_{R_{(aq)}} - \frac{C_{P_{(aq)}}}{k_{eq}} \right) \Delta t \tag{21}$$

Once the volume reaches zero, this interface solid node will be changed to a fluid node. Initialization of information related to fluid flow and mass transport for this new fluid node are implemented using the schemes proposed in Chen et al.[8]. Velocity of this new fluid node is set as zero, as its predecessor is a static solid node; pressure and concentrations of this node are set as the averaged pressure and concentration of its neighboring fluid nodes, respectively.

2.6 Numerical procedures and validations

After initialization, each time stepping involves the following sub-steps: (1) updating the flow field using the fluid flow LB model, then calculating the permeability and porosity; (2) solving the solute transport in the fluid with the dissolution reaction at the fluid-mineral interface using the LB mass transport model, and (3) evolving the mass of solid nodes, updating the geometry of the solid phases and calculating the surface area. Repeating (1)-(3) until the dissolution reaction completes.

The LB pore-scale reactive transport model with dissolution considered has been validated in our previous work [3, 8, 28, 39, 44], and it has been applied to a variety of reactive transport problems [3, 6, 8, 12, 22, 24, 28, 39, 44, 48]. The validation is not repeated here for brevity, and interested readers can refer to our previous work for details.

**3. Results and Discussion**

In this section, effects of mineral heterogeneity on dissolution processes as well as on the evolution of hydrologic properties are investigated. Without loss of generality, a binary mineral system described in 2.1 is considered, which consists of a soluble mineral α, and an insoluble mineral β.

Four types of idealized mineral distributions are first considered as shown in Fig. 1. For all the four cases, the volume fraction of both mineral is 50% in each elementary block shown in the red dashed square. In Case I, α is sandwiched by equal amount of β on the top and bottom, while in Case II β is in the middle while α is outside. Case III and Case IV are obtained by rotating Case I and Case II by 90 degrees, respectively, either clockwise or counter-clockwise. The side length of the elementary block is $a$, and thus the sandwiched mineral height for Cases I and II (or width for Cases III and IV) is $a/2$ while that for outside mineral is $a/4$. Six identical elementary blocks are equidistantly arranged along the flow direction ($x$ direction) in a 400×50 lattices domain, with the gap between subsequent elementary block as $b$. Physical length of each lattice is 10 μm. The centers of the six elementary blocks are located at $y = h/2$, with distance between the inlet and the first elementary block as $w_1$ and that between the outlet and the last elementary block as $w_2$. In the present study, $a = 40$ lattices, $b = 10$ lattices, $h = 50$ lattices, $w_1 = 10$ lattices and $w_2 = 100$ lattices.

It can be seen that Cases I and II have a different orientation compared with Cases III and IV, while the mineral distribution of Cases I and III differs from that of Case II and IV. Therefore, effects of both the mineral orientation and distributions are taken into account by the four seemingly simple structures. After dissolution for these simple structures and distributions is thoroughly investigated under different reactive transport conditions, a more complex mineral distribution case is explored in Section 3.4.

The general physicochemical processes can be described as follows. Initially, mineral α is in chemical equilibrium with $R_{(aq)}$ and $P_{(aq)}$, and thus there is no dissolution reaction. Then reactant $R_{(aq)}$ with a relative high concentration $C_0$ is injected into the domain, causing disequilibrium of the system. Dissolution of α thus takes place, consuming $R_{(aq)}$ and generating $P_{(aq)}$. The undissolved mineral β remains in the system. The boundary and initial conditions are as follows. For fluid flow three kinds of boundary conditions are adopted: a no-slip boundary condition at the fluid-solid interface, periodic boundary conditions for the top and bottom boundaries, and pressure drop applied between the domain inlet and outlet. For solute transport, five types of boundary conditions are used: the concentration boundary condition at the inlet, the outflow boundary condition at the outlet, periodic boundary condition for the top and bottom boundaries, the no-flux boundary condition for β-fluid interface and the reaction boundary condition described by Eq. (5) for the α-fluid interface. For the details of the implementation of these boundary conditions in the LB framework, one can refer to our previous study [24].

The relative strength of convection, diffusion and reaction can be well represented by two important dimensionless numbers, the Peclet number (*Pe*) and Damkohler number (*Da*). *Pe* measures the relative magnitude of convection to diffusion, and *Da* stands for the ratio between reaction and diffusion

$$Pe = \frac{\bar{u}a}{D}, \quad Da = \frac{ka}{D} \tag{24}$$

where $\bar{u}$ is the averaged velocity in the system. Here the definition of *Da* follows that in a series of recent publications [15, 34]. During the dissolution, both the velocity and side length of the elementary lock change, leading to variations of *Pe* and *Da*. In the present study, we discuss the simulation results with the initial *Pe* and *Da*.

It is not necessary to look at all possible combinations of *Da* and *Pe*. Previous studies reported that there are three typical reactive transport conditions according to the relative strength of advection, diffusion and reaction [3, 20, 24] as follows (1) reaction-controlled process, (2) diffusion-controlled process with slow flow and (3) diffusion-controlled process with fast flow. Here, reaction-controlled process means compared with mass transport (either advection or diffusion), the reaction is slow and dissolution is limited by the dissolution reaction rate. Note that for the reaction-controlled case, whether the flow rate is high or low does not make much difference [20]. For the diffusion-controlled process, reaction is fast and diffusion is the limited factor. In such process, however, advection can be strong and quickly supplement the reactant required for dissolution reaction, and this scenario is called diffusion-controlled process with fast flow; the fluid flow can also be negligible and this scenario is called as diffusion-controlled process with slow flow. Wormhole phenomenon appears during diffusion-controlled process with fast flow and face dissolution occurs during diffusion-controlled process with slow flow in porous media with mono-mineral [22]. However, wormhole phenomenon may be suppressed in rocks with undissolved mineral [24]. In the following, the temporal evolutions of the rocks with mineral α and β will be presented and discussed. The variations of important hydrologic properties including porosity, permeability and reactive surface area will be plotted and analyzed.

3.1. Reaction-controlled process

In this scenario, we set $Da = 0.01$, $Pe = 7.4\times10^{-5}$. In the LB framework, the pressure gradient applied is $\Delta p=10^{-5}$, the diffusion coefficient is 0.2 with relaxation time $\tau_g=1.0$, and the dissolution reaction rate as $k_r = 5.0 \times 10^{-5}$. Fig. 2 shows the mineral structures and reactant concentration fields at different times for the four cases studied. Figs. 3(a) and 3(b) display the temporal variations of volume and reactive surface area of mineral α in each elementary block, in which the volume and surface area have been normalized by their initial values, respectively. Note that the six curves from left to right in each image are for elementary blocks 1 to 6 in order, which is also the case in other figures unless other specified. As shown in Fig. 2, due to the fast diffusion compared with reaction, the concentration fields in the domain are rather uniform for all the cases. Correspondingly in each case, the dissolution of mineral α in each elementary block occurs almost simultaneously. However, the dissolution rate of mineral α for different cases differs, and that for Case I and Case III is slower than that of Case II and Case IV, as shown in Fig. 2. This is more evident in Fig. 3. Two important phenomena can be observed in Fig. 3. First, the volume (surface area) variations of Case I and Case III are very close, and that of Case II and Case IV are similar. The time required for complete dissolution of Case I (or Case III) is about $11.25\times10^6$ s, much longer that of Case II (or Case IV). Second, the variations of surface area for Case I (or III) are different from those of Case II (or IV), not only quantitatively but also qualitatively. In Case I and III, the normalized surface area is fixed at unity during almost the entire dissolution process, although with small fluctuations, which drops sharply to zero at the very end, as shown in Fig. 3(b). However, the normalized surface area for Case II and Case IV gradually reduces from unity to zero.

In fact, in the reaction-controlled process, the concentration of reactant $R_{(aq)}$ at the solid-liquid interface is almost uniform in the entire domain (See Fig. 2), so does the concentration of the product $P_{(aq)}$ (concentration field not shown here), due to the fast mass transport. Therefore, the only factor playing a role on the mineral dissolution is the reactive surface area between fluid and mineral α, $S$. Obviously, due to the same distribution, $S$ for Case I and III is the same, while that for Case II and Case IV is identical, and the former one (Case I and III) is smaller than the latter one (Case II and IV), leading to longer time required for the complete dissolution of mineral α for Case I and III. Besides, since mineral α is sandwiched by undissolved mineral β in Case I and III, dissolution can only take place from the two ends, as shown in Figs. 2(a) and 2(c). Thus, during the dissolution in Case I and III, the surface area $S$ does not change, resulting in $S/S_0$ as unity during almost the entire dissolution process.

The differences discussed above for the four cases with different mineral distributions demonstrate that, although the volume fraction of each mineral as well as the overall shape of rocks are the same, the detailed mineral distribution indeed greatly affects the dissolution processes and the hydrologic properties.

3.2. Diffusion-controlled process with slow flow

Now $Da$ number is increased to 5, while $Pe$ is fixed as $7.4\times10^{-5}$ to study the diffusion-controlled process with slow flow. Higher $Da$ is obtained by increasing $k_r$ to $2.5 \times 10^{-2}$ in lattice units, while other parameters are the same as that in Section 3.1. In the diffusion-controlled process with slow flow, diffusion is the dominated mechanism for mass transport, which, however, is slower than the reaction. Thus, face dissolution takes place [24, 39], where dissolution of mineral α only occurs at the α-fluid interface near the inlet at a certain time, as shown in Fig. 4. Due to such face dissolution feature, the elementary blocks from left to right dissolve in sequence, leading to dissolution pattern quite different from the uniform dissolution pattern shown in Fig. 2. The evolutions of volume and surface area are shown in Figs. 5, which are different for different cases and are discussed as follows.

In Case I, for the volume variations shown in Fig. 5(a), there is an inflection point (shown in blue circle) in each curve except the first elementary block, and dissolution rate (indicated by the slope of the volume variation curve) accelerates after the inflection point. Further observation finds that the inflection point in each block corresponds to the time when its left neighboring block is completely dissolved, as shown by the dash red line in Case I in Fig. 5(a). This can be explained as follows. For the mineral distribution in Case I, the complete dissolution of a certain block leads to the following two effects: stopping consuming reactant and generating a wider diffusion path (See $t$=8500s in Fig. 4(a)). The two effects can enhance the mass transport and provide more reactant to the downstream blocks, leading to the accelerated dissolution of downstream blocks. The normalized reactive surface area for each block keeps constant as unity with small fluctuations, which drops sharply to zero at the very end of the dissolution, which is in agreement with the characteristic of face dissolution shown in Fig. 4(a).

For Case II, again the inflection points are observed in the volume variation curve in Fig. 5(a). However in this case, there is hardly any dissolution before the inflection point in each block, compared with that of Case I. This is because reactive surface area of Case II is larger, leading to almost complete consumption of reactant supplied by the slow diffusion from the left inlet, and thus no reactant is available for the downstream block before the dissolution of its neighboring upstream block is finished, as indicated by the concentration fields in Fig. 4(b). The reactive surface shown in Fig. 5(b) for Case II is unity before the inflection point, and it gradually decreases to zero after the inflection point, in agreement with the dissolution pattern in Fig. 4(b).

Case III has the same initial reactive surface area as Case I. While in Section 3.1 (see Fig. 3) temporal evolutions of volume and surface area for Case I and III are quite close, they differ significantly from each other in Fig. 5. Dissolution in Case III is much slower than that in Case I. In fact compared with Case I, there are opposite factors affecting the dissolution in Case III. On one hand, the reactive surface area for each block in Case III is greater than its initial value for most of the time, as indicated by the inclined α-fluid interface due to the face-dissolution feature in Fig. 4(c). This can be seen more clearly in Fig. 5(b) where an $S/S_0$ peak of about 1.4 is obtained. The increased reactive surface area certainly can facilitate the dissolution. On the other hand, mass transport resistance in Case III is higher than Case I, as reactant has to diffuse deep into the vertical gap, which slows down the dissolution in Case III. The extremely slow diffusion mass transport considered here renders the latter adverse factor to play the dominant role, resulting in a longer time required for the complete dissolution for Case III than that for Case I.

Finally, for Case IV, its dissolution rate is slower than Case II, although both cases have the same initial reactive surface area. This is also due to the higher mass transport resistance in Case IV. For the surface area curve in Fig. 5(b), it is interesting that there is an inflection point at about $S/S_0=0.5$. This inflection point indicates the complete dissolution of the left half α in each elementary block (See Fig. 4(d)). Obviously in each block its left half α dissolves faster than its right half, due to more reactant there (see Fig. 4(d)). Actually, the fact that the inflection point occurs at $S/S_0$=0.5 means the right half is hardly dissolved before the left half is completely dissolved away. The dissolution rate reduces after the inflection points, as indicated by the slope reduction of the curves after the inflection points in Fig. 5 for Case IV.

From the above discussion, it can be seen that for the diffusion-controlled process with slow flow, the variations of volume and reactive surface are more complicated than the reaction-controlled process in Section 3.1. Although initial rock structures or even the reactive surface areas are the same, dissolution processes can be quite different due to different mineral distributions. In fact, "effective reactive surface area" should be distinguished from "reactive surface area" [15, 34, 36]. For the fast mass transport case such as that in Section 3.1, they are the same. However, when mass transport rate becomes a limited factor, only at those surfaces where reactant can be efficiently supplemented the reaction is allowed, and thus the corresponding reactive surface areas are effective. For the reactive transport processes studied in Fig. 4 with face-dissolution, only those surface facing the inlet is effective. Therefore, although Case III has the same initial reactive surface area as Case I, because reactant has to diffuse deep into the vertical gap to reach the reactive surface area in Case III, the "effective reaction surface area" for Case III is smaller than Case I, thus leading to slower dissolution in Case III. In the same way, although reactive surface areas for Case I and II are different, their effective surface areas are almost the same (Figs. 5(a) and (b)), resulting in similar time required for complete dissolution.

3.3 Diffusion-controlled process with fast flow

In this section, reactive transport process of diffusion-controlled process with fast flow is considered. $Da$ is set as 5, the same as that in Section 3.2, while $Pe$ is increased to 18.6 by increasing the pressure gradient between inlet and outlet from $10^{-5}$ to 0.05 (lattice unit) and reducing diffusion coefficient from 0.2 to 0.04 with relaxation time $\tau_g$=0.51.

Fig. 6 shows time evolutions of the mineral structures and reactant concentration fields for the four cases. As shown in Fig. 6, advection in the channel (half on the top and half on the bottom, indicated by the arrows in Fig. 6(a), called as "initial channel" hereinafter) is strong, and reactant mainly transports in this channel while void space between subsequent blocks is bypassed. Such transport process is the typical feature in diffusion-controlled process with fast flow [3, 24]. Therefore, for the four cases, it is expected that the case with the highest reactive surface area adjacent to the initial channel has the fastest dissolution rate. Consequently, the gross dissolution rate of Case II is the quickest, followed by Case IV, III and lastly Case I, as shown in Fig. 7. A close observation of the curves for Case II in Fig. 7 finds that the difference of time required for complete dissolution between subsequent blocks decreases from left to right. In other words, the time difference for complete dissolution between blocks 1 and 2 is the highest, and that between blocks 5 and 6 is the lowest. This is because the dissolution widens the flow channel in turn providing more reactant to the flow channel (Fig. 6(b)), leading to the positive feedback between mass transport and reaction which is well known as the wormhole phenomenon [3, 24]. However, such positive feedback is suppressed in Case I, Case III and Case IV due to the undissolved mineral which distributes adjacent to the preferred channel. This suppressing effects are also found in our previous studies [24]. For Case IV, although its initial reactive surface area is the same as Case II, its effective reactive surface is lower because these reactive surface areas are located vertically in the gaps between neighboring blocks, which are bypassed by the reactant transport, leading to locally slow diffusion-dominated transport mechanism (See Fig. 6(d)). Dissolution in Case III is expected to be slower than in Case IV for two reasons. First, its reactive surface area is smaller than Case IV, and second, diffusion resistance in the bypassed void space is higher. Finally, for Case I, because there is no reactive surface directly adjacent to the preferred flow channel, its dissolution rate is the slowest. An interesting phenomenon for Case I is that, the final block (Block 6) dissolves faster than blocks 2-5, while in all other scenarios studied in the present study the six block dissolves subsequently, as shown in Fig. 7. This is because the reactant arriving at the outlet is still with high concentration, because the consumption through the domain is slight as no reactive surface area is directly adjacent to the preferred channel; near the outlet, the reactant with high concentration diffuses back to the right surface of the last block ($t = 600$s in Fig. 6(a)), causing the surface of last block facing the outlet to dissolve. The dissolution rate at this surface is faster compared with other blocks (Fig. 6(a)). Such "tail effect" also can be observed in the last block for Case IV in Fig. 6(d), where the right half α dissolves relatively quicker than its corresponding left half (t=360s in Fig. 6(d)). However, as the diffusion resistance in the gap between subsequent blocks in Case IV is not as significant as that in Case I, the six block still dissolves subsequently (see Case IV in Fig. 7). From the above discussion, it can be found that the completion between convection, diffusion and reaction, along with the heterogeneous mineral distributions, make the interaction between structure, reactive transport and dissolution even more complicated.

3.4 Relation between reactive surface area and volume

Results and discussions in Sections 3.1 to 3.3 reveal that the reactive surface area plays an important role on the reactive transport processes. While there have been some techniques (vertical scanning interferometry, atomic force microscopy, laser confocal microscopy, X-ray tomography) for measuring the reactive surface area [36, 37], it is still challenging to directly measure the reactive surface area. In the literature, several models based on geometrical constructions have been proposed to describe the relationship between the reactive surface area and mineral volume. The typical models are spherical-grain model and spherical-pore model [36]. The spherical-grain model assumes that the porous medium is composed of suspending spherical grains and the reactive surface area reduces with volume according to $S/S_0 = (\frac{V}{V_0})^{2/3}$, while in the spherical-pore model the pores are suspending spheres and the reactive surface area grows up as volume decreases following $S/S_0 = (\frac{V}{V_0})^{-2/3}$. These two models give opposite tendencies of the reactive surface area relative to volume, indicating the complexity of the relation.

Fig. 8 displays the relationships between the reactive surface area and volume of mineral α for the four cases studied under different reactive transport processes. As shown in Fig. 8, the relationships vary for different scenarios. Besides, while a few of them can be approximately described by the spherical-grain model, most of them show large discrepancy with the spherical-grain model, not only quantitatively but also qualitatively, again demonstrating the complexity of the effects of mineral distributions and compositions on the dissolution processes.

Fig. 8(a) is for the reaction-controlled processes studied in Section 3.1. It can be seen that the evolutions of $S/S_0$ with volume $V/V_0$ for Cases I and III are similar, and those for Case II and IV coincide. For Cases I and III, the reactive surface area keeps constant which falls down only when the volume is close to zero indicating the end of the dissolution, while for Cases II and IV, $S/S_0$ continuously declines as volume reduces. These observations are consistent with the discussions in Section 3.1. Cases II and IV can be roughly described by the spherical-grain model, but using the spherical-grain model for Case I and III will lead to large error (note that the spherical-grain model in two-dimension simulation is given by $S/S_0 = (\frac{V}{V_0})^{1/2}$ instead of $S/S_0 = (\frac{V}{V_0})^{2/3}$).

Fig. 8(b) displays $S/S_0$ vs. $V/V_0$ for diffusion controlled process with slow flow studied in Section 3.2. For Cases II and IV, $S/S_0$ almost linearly decreases as volume reduces, due to their mineral distribution characteristics and the face dissolution process. The curves for Cases I and III, drop down stepwise, with each step and the sharp drop at the end of each step representing the dissolution proceeding and dissolution end of each elementary block, agreeing with the discussions in Section 3.2. The relationships between $S/S_0$ and $V/V_0$ for all the cases deviate from the spherical-grain model, not only quantitatively but also qualitatively.

Fig. 8(c) shows $S/S_0$ vs. $V/V_0$ for diffusion controlled process with fast flow studied in Section 3.3. The curves for Cases II and IV are quite close, both of which can be well described by the spherical-grain model. The curves for Cases I and III, roughly the same, but differ from each other locally. Although the curves for Case I and III still can be described as stepwise, the length of each step is not uniform, with the first step relatively long. This is because for the dissolution process with fast flow in Section 3.3, all the elementary blocks dissolve during the first step, with the sharp drop in the first step standing for the complete dissolution of the first block, while in Fig. 8(b) only one elementary block dissolves in each step caused by the face-dissolution. Besides, the drop of the first step for Case III is delayed compared with that for Case I, which is because mineral α in Block 1 in Case I dissolves faster than that in Case III as the left face of the former case directly contacts with the inlet.

Comparing the curves in Fig. 8, it can be found that the relationship between $S/S_0$ and $V/V_0$ is not only affected by the mineral distributions, but also is a function of the particular reactive transport condition. Even for the relatively simple structure and distribution studied, various relationships are observed, including constant $S/S_0$ during the entire dissolution processes with sharp drop at the very end, stepwise pattern, linear function, and that well described by the spherical-grain model. Therefore, using one specific model to determine the reactive surface area during dissolution processes for more complex structures and mineral compositions in continuum-scale model will definitely generate large errors, and would predict inaccurate or even wrong physicochemical behaviors. Unfortunately, based on the results in Fig. 8, it seems that it is extremely challenging, if possible, to propose a generalized relationship that can be adopted in the continuum-scale model, due to very complex interplay of the porous structures, mineral compositions as well as the reactive transport processes [36].

3.4 Porosity and permeability

Permeability is an important transport property of a porous medium, which is an indicator of the capacity for fluid flowing through the porous medium. It is also an important input parameter for continuum-scale simulations and greatly affects the predicted large-scale flow behaviors. During our simulations, the mineral dissolution causes the structure change, and thus both the porosity and permeability alter during the dissolution. The dynamic evolutions of porosity and permeability are monitored and are plotted in Fig. 9. Note that both the porosity and permeability are normalized using their initial values, which are the same for all the four cases as 52.1% and $2.48\times10^{-10}$ m$^2$, respectively. As can be seen from Fig. 9, the porosity-permeability relationships for different cases under different $Da$ and $Pe$ generally differ from each other, indicating that both the mineral distributions and the reactive transport conditions play important roles on the porosity-permeability relationships. It is obvious that the final normalized porosity should be the same for all the cases, which is 1.46, as shown in Fig. 9. The final normalized permeability, however, is different for cases, with Case II having the highest permeability of 24.6, followed by Case I, Case IV and finally Case III with the lowest permeability of 1.43.

For the same reactive transport condition (the same $Pe$ and $Da$), the porosity-permeability curves are quite different for different cases. As shown in Fig. 9(a) corresponding to the reaction-controlled process with uniform dissolution, for Case I the permeability remains constant until $\varepsilon/\varepsilon_0$ is greater than 1.42. This is because for Case I, the middle channel between the undissolved mineral β will not be opened until the complete dissolution of mineral α, before which the fluid mainly flows in the initial channel. For Case II, with $k/k_0$ plotted as half-log scale, it can be seen that $\log(k/k_0)$ approximately linearly increases as $\varepsilon/\varepsilon_0$, indicating a porosity-permeability relationship as $\log(k/k_0) \propto \varepsilon/\varepsilon_0$. Case III and IV show the similar change trend, and the rate of increase of $k/k_0$ gradually decreases as dissolution proceeds. This is because for these two cases, as dissolution proceeds, the vertical void space between undissolved mineral β contributes less to the fluid flow in the initial channel.

Fig. 9(b) presents the $k/k_0$-$\varepsilon/\varepsilon_0$ relationship for diffusion-controlled process with slow fluid flow. For Case I, the permeability grows up stepwise with porosity, with each step indicating the complete dissolution of mineral α in each elementary block which opens the channel between the undissolved mineral β. Case II, in which the permeability raises with porosity at an increasing rate, shows the largest permeability all the time in the four cases. Compared with Fig. 9(a), it can be found that the curve for Case II is concave rather than linear. For Cases III and IV, the permeability increases gradually when the face-dissolution occurs from left inlet to right outlet, which is different from that in Fig. 9(a) where the increase of $k/k_0$ ceases at a relatively lower $\varepsilon/\varepsilon_0$, for example $\varepsilon/\varepsilon_0=1.25$ for Case III in Fig. 9(a).

Fig. 9(c) plots the $k/k_0$-$\varepsilon/\varepsilon_0$ relationship for diffusion-controlled process with fast flow. Although the evolutions of $S/S_0$ and $V/V_0$ as well as the $S$–$V$ relationship for this process is quite different from the reaction-controlled process (see Figs. 3 and 7, Figs. 8(a) and 8(c)), the porosity-permeability relationships in this reaction transport condition is close to that for the reaction-controlled process. This is because the dissolution in diffusion-controlled process with fast flow can roughly be described by uniform dissolution due to very fast flow studied, as shown in Fig. 6 and Fig. 7 (the curves for all the cases are close). However, there are some local differences for two reactive transport conditions. For example for Case I, the $k/k_0=1$ regions ceases at a lower $\varepsilon/\varepsilon_0$ in Fig. 9(c) than that in Fig. 9(a), due to the fact that for diffusion-controlled process with fast flow, open of gaps in each elementary block are not simultaneous (as shown in Fig. 6(a)).

From the above discussions, it can be found that the porosity-permeability relationship is a complex function of mineral distributions and reaction transport conditions, and it is also challenging to draw a generalized relationship. Therefore, using single porosity-permeability relationship to predict fluid flow in continuum-scale modeling also would lead to inaccurate fluid flow behaviors.

3.5 Dissolution of rocks with random-distributed undissolved mineral

Now a relatively complex mineral distribution is studied. Particles of mineral β with square shape of size 4×4 lattices are randomly dispersed in each elementary block. Several distributions are generated, and the one in which all the mineral α is connected is selected for simulation, as shown in Fig. 10. In other words, all α in Fig. 10 can be dissolved. Other structural parameters are the same as that in Section 4.1. The three typical reactive transport processes are also studied. The evolutions of reactant concentration and rock structures for different reactive transport processes are shown in Figs. 10(a), 10(c) and 10 (e), and the corresponding variations of $S/S_0$ and $V/V_0$ are shown at the right as Fig. 10(b), 10 (d) and 10(f). Generally, the dissolution characteristics, the time evolutions of $S/S_0$ and $V/V_0$ are similar to that discussed in Sections 3.1-3.3, thus are not repeated here for brevity. Note that porous structures are formed by the remaining undissolved mineral β, in which mass transport resistance is extremely high, leading to longer time required for complete dissolution of mineral α in Fig. 10 compared with that in Sections 3.1-3.3. Besides, the variations of the reactive surface area are more complex due to the effects of structure heterogeneity. The relationship between $S/S_0$ and $V/V_0$ is shown in Fig. 11(a), and it can be found that none of the curves can be well described by the spherical model. Besides, the relationship is affected by the reactive transport conditions. Fig. 11(b) further plots the variations of $k/k_0$ with $\varepsilon/\varepsilon_0$. The reaction-controlled process and the diffusion-controlled process with fast flow show similar trends, for which permeability increases relatively fast first and then reaches a constant value at a relatively lower porosity about $\varepsilon/\varepsilon_0=1.1$. This is because dissolution deep in the elementary block contributes less to the enhancement of fluid flow [24]. For the diffusion-controlled process with slow flow, $k/k_0$ gradually increases in the entire dissolution process, due to the face-dissolution feature. Again, the reactive transport conditions greatly affect the permeability-porosity relationship.

## 4 Conclusion

In this work heterogeneities of mineral distributions and mineral dissolution reaction rate are studied. Dissolution in rocks with binary minerals is investigated with mineralogical heterogeneity considered. A LB pore-scale reactive transport model is adopted to simulate the coupled fluid flow, species transport, reaction and solid dissolution processes. A relatively simple rock structure with idealized mineral distributions is employed for the purpose to understand in more depth the reactive transport phenomena and to reveal the coupling mechanisms between advection, diffusion, reaction as well as the mineralogical heterogeneity. Three typical reactive transport processes including reaction-controlled, diffusion-controlled with slow flow and diffusion-controlled with fast flow processes are explored. Simulation results show significant effects of mineral distribution and mineral chemical heterogeneity on the dissolution process. Although general uniform, face and channel dissolution features are maintained at the domain scale under the three typical reactive transport processes, the existence of undissolved mineral causes complex local dissolution behaviors. Important hydrologic properties including porosity, permeability and reactive surface area are monitored. It is found that the mineral distribution and chemical heterogeneity also greatly affect the evolutions of these hydrologic properties. Change trends of the hydrologic properties usually differ significantly for different mineral distributions and different reactive transport conditions. Various surface area-volume relationships are obtained, including constant surface area during nearly the entire dissolution processes, stepwise change, linear function, and that can be well described by the spherical-grain model. The porosity-permeability relationships for different mineral distributions and different reactive transport conditions are also different. The present study indicates that using one specific model or single relationship to determine changes of reactive surface area and permeability in the continuum-scale models would generate large errors and predict inaccurate or even wrong physicochemical behaviors. Based on the simulation results in the present study, it is found that the mineralogical heterogeneity further complicates the scaling effect existing between continuum scale and pore scale. Further, it seems that it is extremely challenging, if possible, to propose a generalize relationship to calculate the reactive surface area and permeability during dissolution, due to complex coupling and strong interactions between the hydrodynamic conditions, porous structure heterogeneity as well as the mineralogical heterogeneity.

**Acknowledgement**

The authors also thank the support of National Nature Science Foundation of China (51406145 and 51136004). The authors acknowledge the support of LANL's LDRD Program and Institutional Computing Program.

**Reference**

1. Steefel, C.I., D.J. DePaolo, and P.C. Lichtner, *Reactive transport modeling: An essential tool and a new research approach for the Earth sciences.* Earth and Planetary Science Letters, 2005. **240**(3): p. 539-558.
2. Lichtner, P.C., *Continuum formulation of multicomponent-multiphase reactive transport.* Reviews in Mineralogy and Geochemistry, 1996. **34**(1): p. 1-81.

3.  Kang, Q., et al., *Lattice Boltzmann simulation of chemical dissolution in porous media.* Physical Review E, 2002. **65**(3): p. 036318.
4.  Li, L., C.A. Peters, and M.A. Celia, *Upscaling geochemical reaction rates using pore-scale network modeling.* Advances in Water Resources, 2006. **29**(9): p. 1351-1370.
5.  Evans, R.L. and D. Lizarralde, *Geophysical evidence for karst formation associated with offshore groundwater transport: an example from North Carolina.* Geochemistry, Geophysics, Geosystems, 2003. **4**(8).
6.  Chen, L., et al., *Mesoscopic study of the effects of gel concentration and materials on the formation of precipitation patterns.* Langmuir, 2012. **28**(32): p. 11745-11754.
7.  Cubillas, P., et al., *How do mineral coatings affect dissolution rates? An experimental study of coupled $CaCO_3$ dissolution—$CdCO_3$ precipitation.* Geochimica et Cosmochimica Acta, 2005. **69**(23): p. 5459-5476.
8.  Chen, L., et al., *Pore-scale modeling of multiphase reactive transport with phase transitions and dissolution-precipitation processes in closed systems.* Physical Review E, 2013. **87**(4): p. 043306.
9.  Luquot, L. and P. Gouze, *Experimental determination of porosity and permeability changes induced by injection of $CO_2$ into carbonate rocks.* Chemical Geology, 2009. **265**(1–2): p. 148-159.
10. Stockmann, G.J., et al., *Do carbonate precipitates affect dissolution kinetics?: 2: Diopside.* Chemical Geology, 2013. **337**: p. 56-66.
11. Sanna, A., et al., *A review of mineral carbonation technologies to sequester $CO_2$.* Chemical Society Reviews, 2014. **43**(23): p. 8049-8080.
12. Chen, L., et al., *Pore-scale simulation of multicomponent multiphase reactive transport with dissolution and precipitation.* International Journal of Heat and Mass Transfer, 2015. **85**: p. 935-949.
13. Waples, D.W., *Geochemistry in petroleum exploration* 2013: Springer Science & Business Media.
14. Molins, S., et al., *An investigation of the effect of pore scale flow on average geochemical reaction rates using direct numerical simulation.* Water Resour. Res, 2012. **48**(3): p. W03527.

15. Tartakovsky, A.M., et al., *A smoothed particle hydrodynamics model for reactive transport and mineral precipitation in porous and fractured porous media.* Water Resour. Res, 2007. **43**(5): p. W05437.
16. Liu, Y., et al., *Pore and continuum scale study of the effect of subgrid transport heterogeneity on redox reaction rates.* Geochimica et Cosmochimica Acta, 2015. **163**: p. 140-155.
17. Liu, C., et al., *Pore-scale process coupling and effective surface reaction rates in heterogeneous subsurface materials.* Rev Mineral Geochem, 2015. **80**: p. 191-216.
18. Ovaysi, S. and M. Piri, *Pore-space alteration induced by brine acidification in subsurface geologic formations.* Water Resources Research, 2014. **50**(1): p. 440-452.
19. Szymczak, P. and A.J.C. Ladd, *Wormhole formation in dissolving fractures.* Journal of Geophysical Research: Solid Earth, 2009. **114**(B6): p. n/a-n/a.
20. Bekri, S., J. Thovert, and P. Adler, *Dissolution of porous media.* Chemical Engineering Science, 1995. **50**(17): p. 2765-2791.
21. Huber, C., B. Shafei, and A. Parmigiani, *A new pore-scale model for linear and non-linear heterogeneous dissolution and precipitation.* Geochimica et Cosmochimica Acta, 2014. **124**: p. 109-130.
22. Kang, Q., et al., *Pore-scale study of dissolution-induced changes in permeability and porosity of porous media.* Journal of Hydrology, 2014. **517**: p. 1049-1055.
23. Ryan, E.M., A.M. Tartakovsky, and C. Amon, *Pore-scale modeling of competitive adsorption in porous media.* Journal of Contaminant Hydrology, 2011. **120–121**: p. 56-78.
24. Chen, L., et al., *Pore-scale study of dissolution-induced changes in hydrologic properties of rocks with binary minerals.* Water Resources Research, 2014. **50**(12): p. 9343-9365.
25. Noiriel, C., B. Madé, and P. Gouze, *Impact of coating development on the hydraulic and transport properties in argillaceous limestone fracture.* Water resources research, 2007. **43**(9).
26. Noiriel, C., P. Gouze, and B. Madé, *3D analysis of geometry and flow changes in a limestone fracture during dissolution.* Journal of hydrology, 2013. **486**: p. 211-223.
27. Li, L., C.A. Peters, and M.A. Celia, *Effects of mineral spatial distribution on reaction rates in porous media.* Water Resources Research, 2007. **43**(1): p. n/a-n/a.

28. Kang, Q., P.C. Lichtner, and D. Zhang, *Lattice Boltzmann pore-scale model for multicomponent reactive transport in porous media.* Journal of Geophysical Research: Solid Earth (1978–2012), 2006. **111**(B5).
29. Yoon, H., Q. Kang, and A.J. Valocchi, *Lattice Boltzmann-based approaches for pore-scale reactive transport.* Reviews in Mineralogy & Geochemistry, 2015. **80**: p. 393-431.
30. Chen, L.-Q., *Phase-filed models for microstructures evolution.* Annu. Rev. Mater. Res., 2002. **32**: p. 113–140.
31. Sun, D., et al., *Lattice Boltzmann modeling of dendritic growth in a forced melt convection.* Acta Mater, 2009. **57**(6): p. 1755-1767.
32. Li, X., H. Huang, and P. Meakin, *Level set simulation of coupled advection-diffusion and pore structure evolution due to mineral precipitation in porous media.* Water Resour. Res., 2008. **44**(12): p. W12407.
33. Vu, M.T. and P.M. Adler, *Application of level-set method for deposition in three-dimensional reconstructed porous media.* Physical Review E, 2014. **89**(5): p. 053301.
34. Lichtner, P.C. and Q. Kang, *Upscaling pore-scale reactive transport equations using a multiscale continuum formulation.* Water Resources Research, 2007. **43**(12): p. n/a-n/a.
35. Sadhukhan, S., P. Gouze, and T. Dutta, *A simulation study of reactive flow in 2-D involving dissolution and precipitation in sedimentary rocks.* Journal of Hydrology, 2014. **519, Part B**: p. 2101-2110.
36. Noiriel, C., et al., *Changes in reactive surface area during limestone dissolution: An experimental and modelling study.* Chemical Geology, 2009. **265**(1–2): p. 160-170.
37. Lai, P., K. Moulton, and S. Krevor, *Pore-scale heterogeneity in the mineral distribution and reactive surface area of porous rocks.* Chemical Geology, 2015. **411**: p. 260-273.
38. Kang, Q., D. Zhang, and S. Chen, *Unified lattice Boltzmann method for flow in multiscale porous media.* Physical Review E, 2002. **66**(5): p. 056307.
39. Kang, Q., D. Zhang, and S. Chen, *Simulation of dissolution and precipitation in porous media.* Journal of Geophysical Research: Solid Earth (1978–2012), 2003. **108**(B10).
40. Zhang, J., *Lattice Boltzmann method for microfluidics: models and applications.* Microfluidics and Nanofluidics, 2011. **10**(1): p. 1-28.
41. Chen, S. and G. Doolen, *Lattice Boltzmann method for fluid flows.* Annual review of fluid mechanics, 1998. **30**: p. 329-364.

42. Chen, L., et al., *A critical review of the pseudopotential multiphase lattice Boltzmann model: Methods and applications.* International Journal of Heat and Mass Transfer, 2014. **76**: p. 210-236.
43. Guo, Z., B. Shi, and N. Wang, *Lattice BGK model for incompressible Navier–Stokes equation.* Journal of Computational Physics, 2000. **165**(1): p. 288-306.
44. Kang, Q., P.C. Lichtner, and D. Zhang, *An improved lattice Boltzmann model for multicomponent reactive transport in porous media at the pore scale.* Water Resources Research, 2007. **43**(12).
45. Zhang, T., et al., *General bounce-back scheme for concentration boundary condition in the lattice-Boltzmann method.* Physical Review E, 2012. **85**(1): p. 016701.
46. Gillissen, J.J.J. and N. Looije, *Boundary conditions for surface reactions in lattice Boltzmann simulations.* Physical Review E, 2014. **89**(6): p. 063307.
47. Walsh, S.D.C. and M.O. Saar, *Interpolated lattice Boltzmann boundary conditions for surface reaction kinetics.* Physical Review E, 2010. **82**(6): p. 066703.
48. Chen, L., et al., *Coupled numerical approach combining finite volume and lattice Boltzmann methods for multi-scale multi-physicochemical processes.* Journal of Computational Physics, 2013. **255**: p. 83-105.

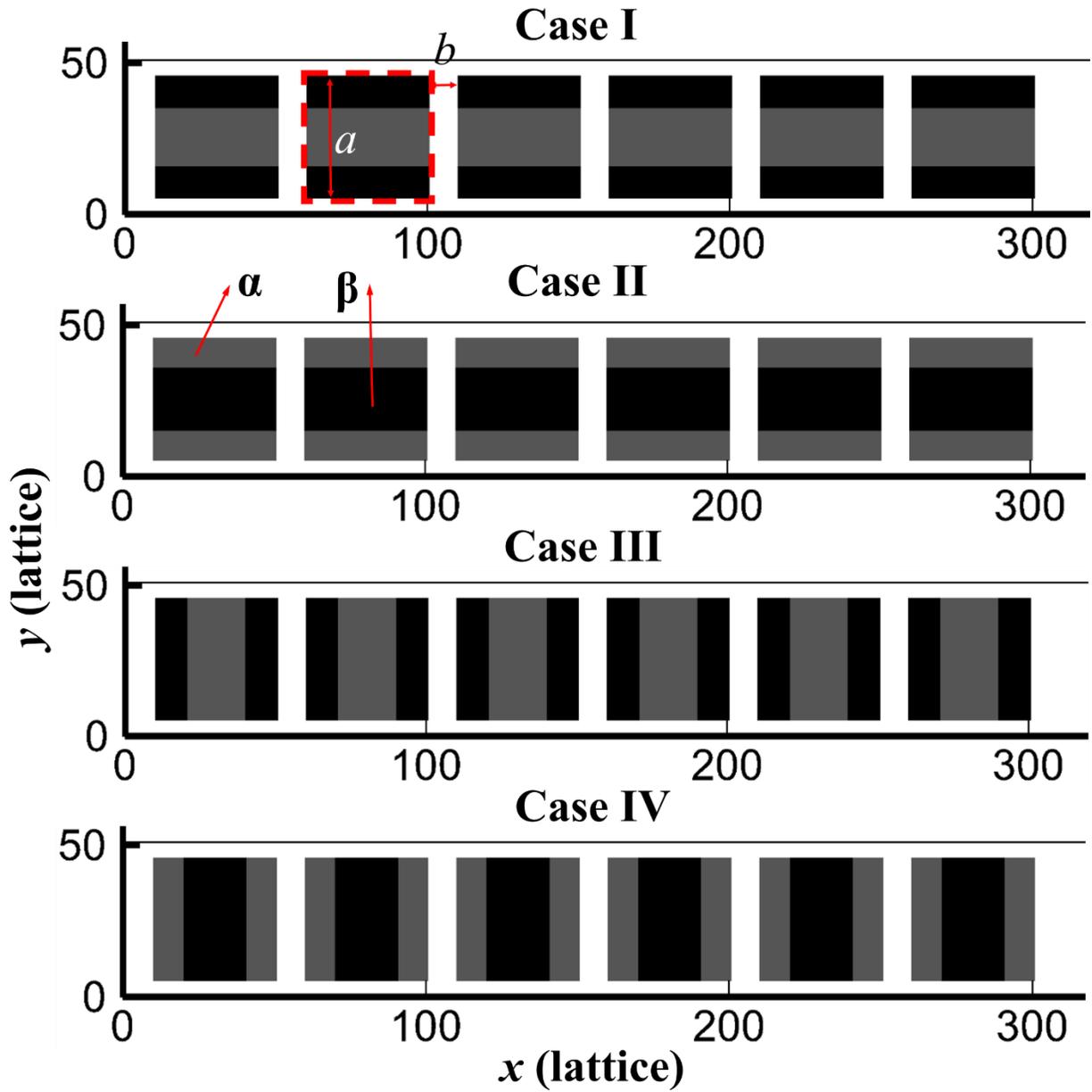

Fig. 1 Rocks with four idealized mineral distributions. The gray and black regions are mineral α and mineral β, respectively. In the red dashed square locates an elementary block.

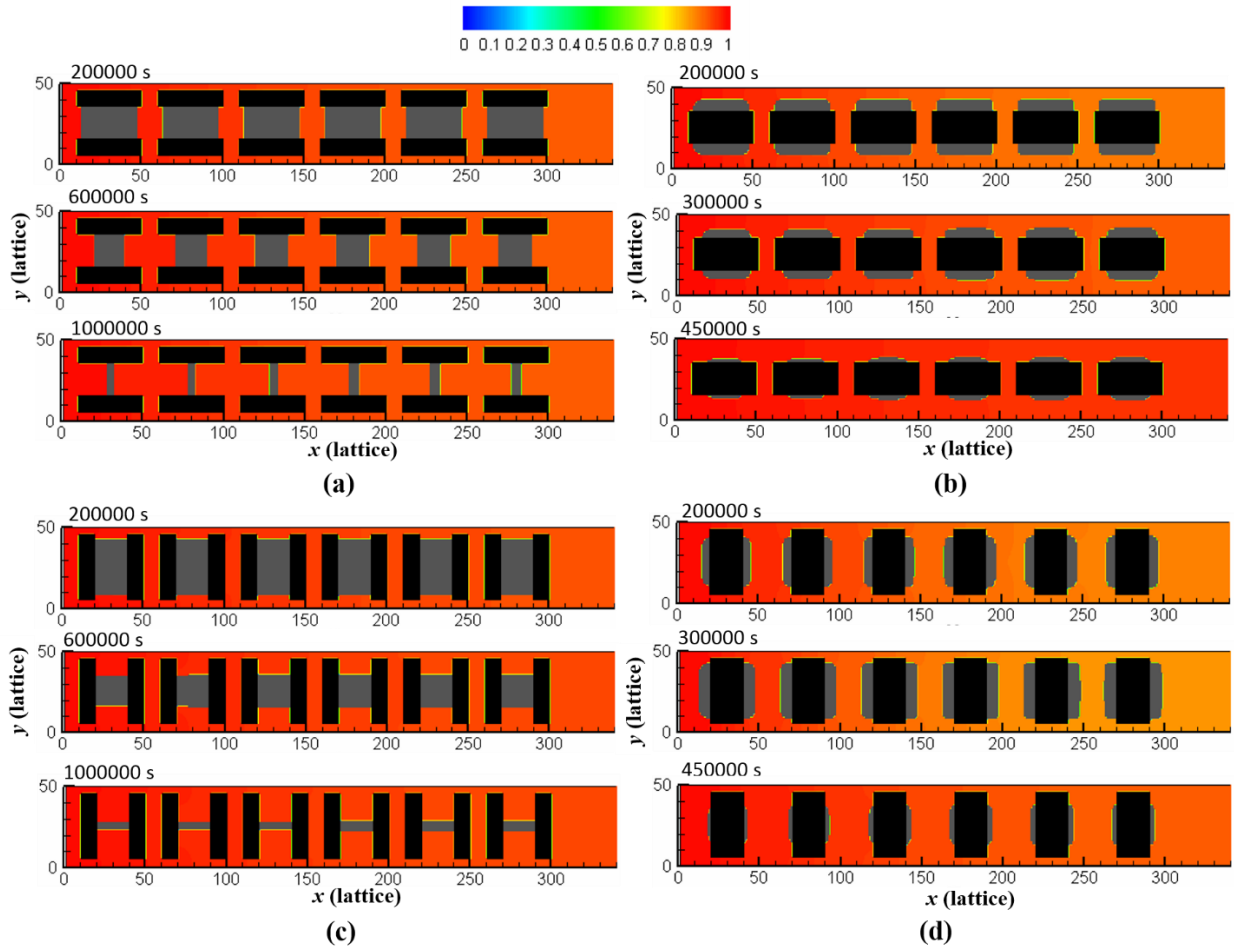

Fig. 2 Reactant concentration distribution and morphology of minerals at the beginning, middle and the last stages during reaction-controlled processes with $Pe = 7.4\times10^{-5}$ and $Da = 0.01$. The gray region denotes soluble mineral α, the black zone represents insoluble mineral β. Reactant concentration varies from 1.0 M to 0.0 M when the legend changes from red to blue. (a) Case I, (b) II, (c) Case III, (d) Case IV.

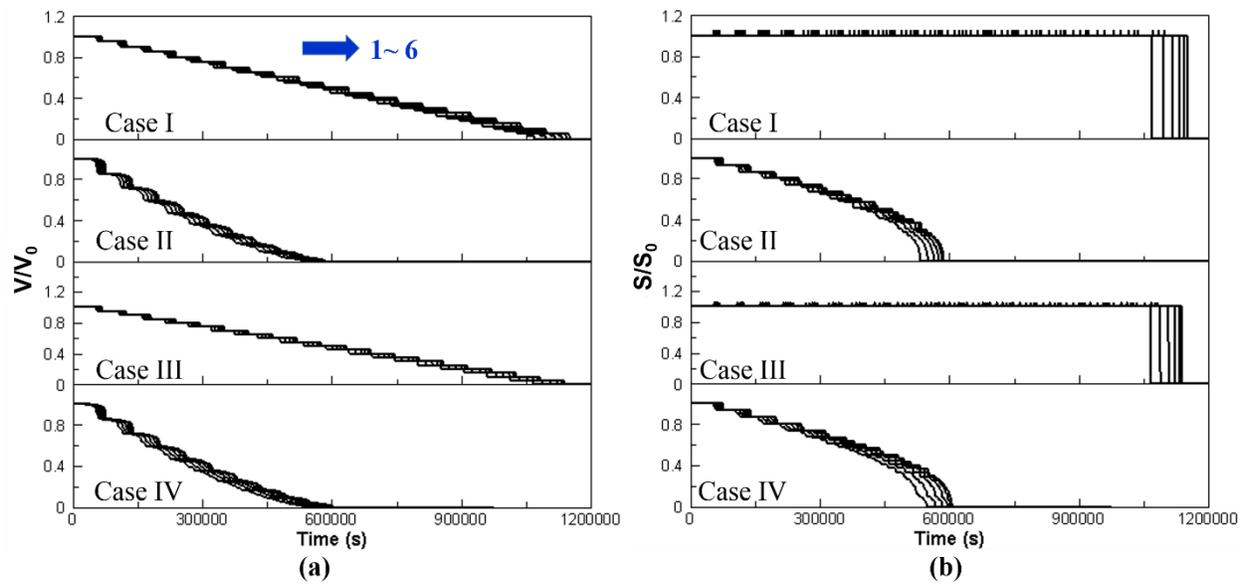

Fig 3 Evolution of volume (a) and surface area (b) of mineral α in each elementary block during reaction-controlled process. In each image, from top to bottom, Case I, Case II, Case III and Case IV.

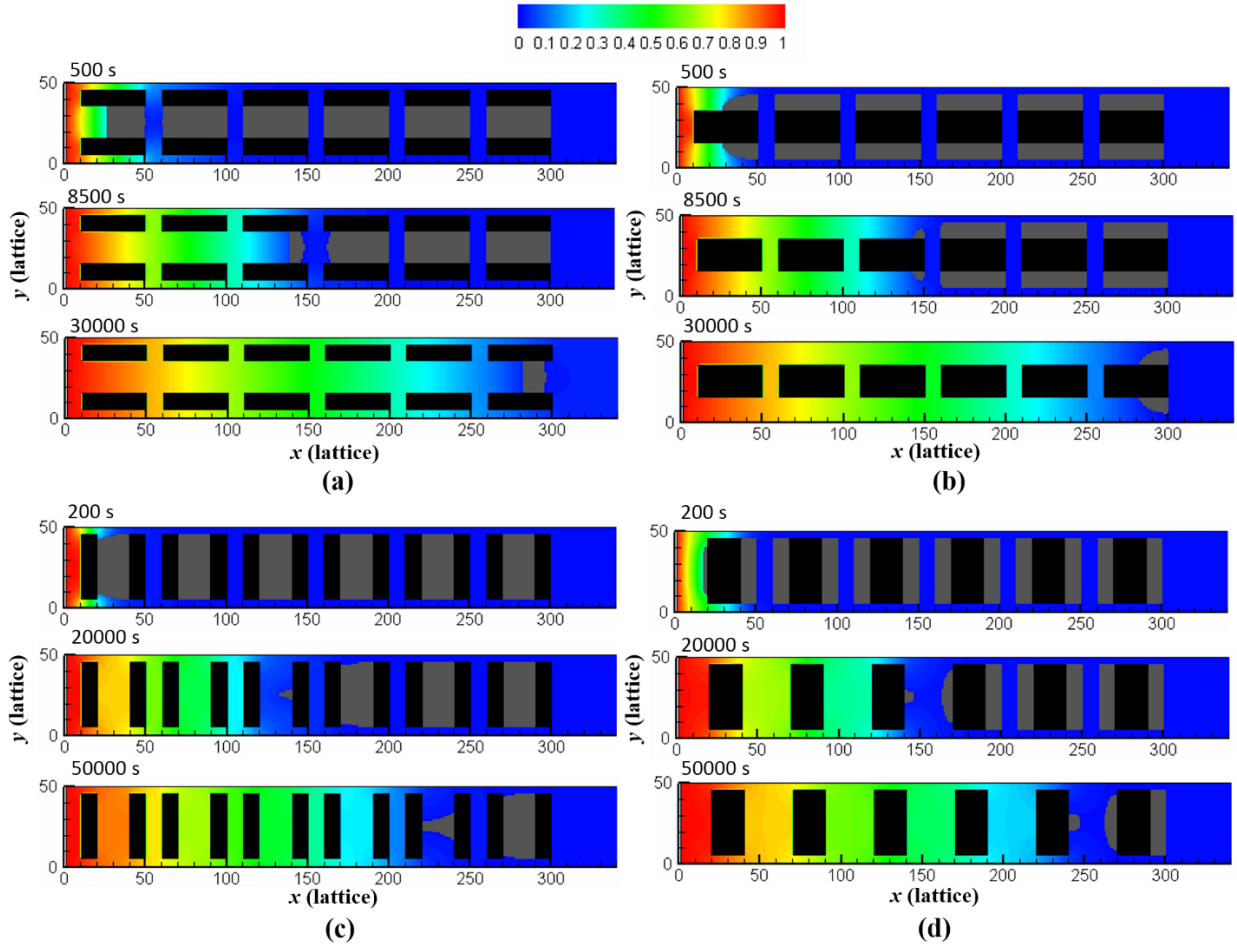

Fig.4 Reactant concentration distribution and morphology of minerals at the beginning, middle and the last stages during diffusion-controlled processes with slow flow where $Pe = 7.4\times10^{-5}$ and $Da = 5$. (a) Case I, (b) II, (c) Case III, (d) Case IV.

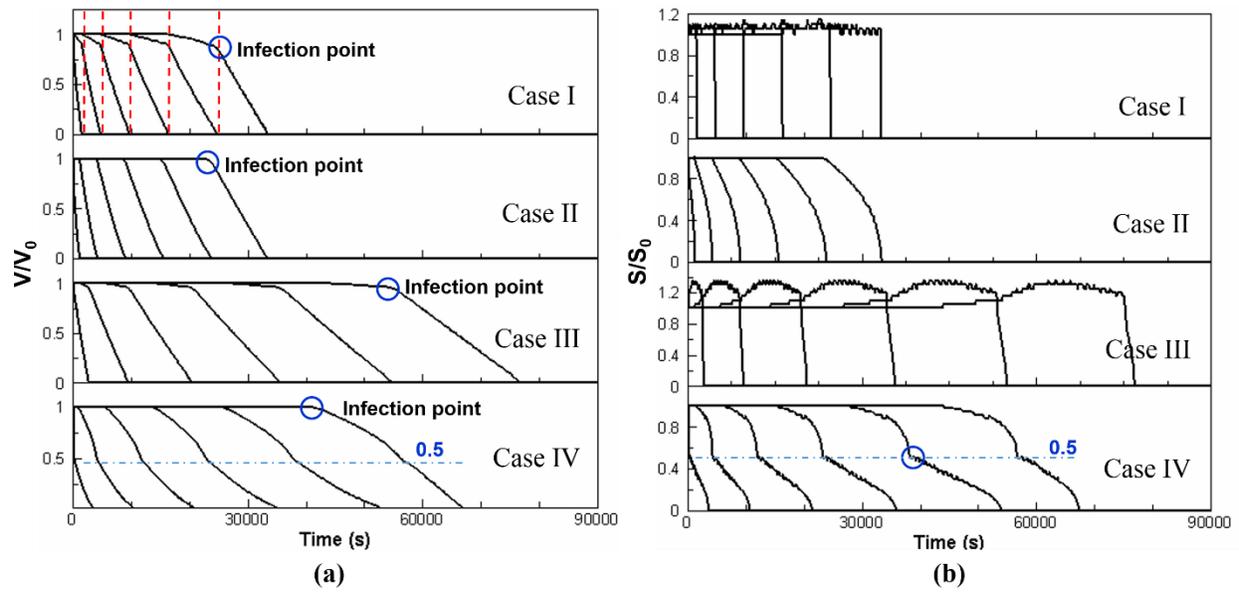

Fig 5 Evolution of volume (a) and surface area (b) of mineral α in each elementary block during diffusion-controlled process with slow flow. In each image, from top to bottom, Case I, Case II, Case III and Case IV.

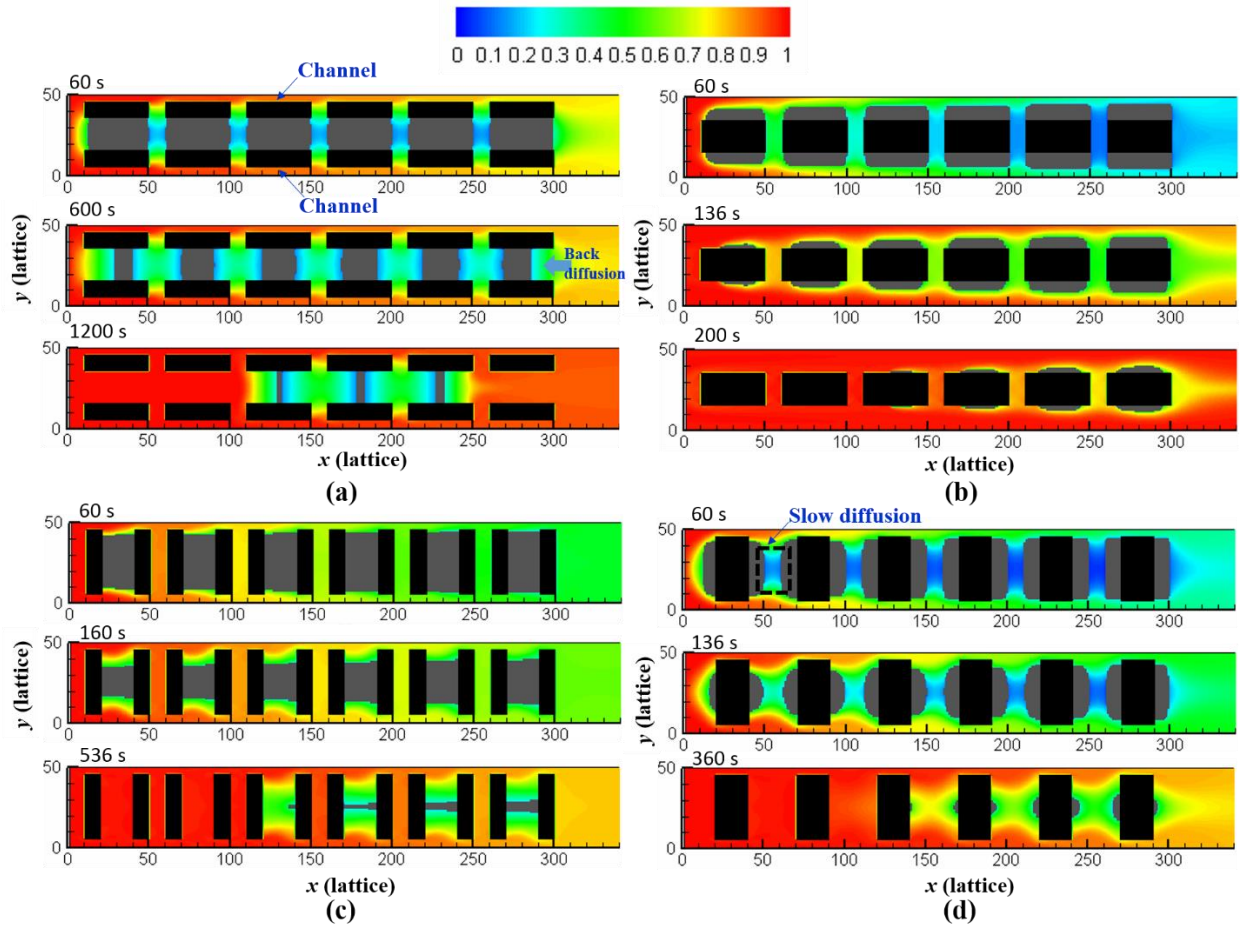

Fig.6 Reactant concentration distribution and morphology of minerals at the beginning, middle and the last stages during diffusion-controlled processes with fast flow where $Pe$ = 18.6 and $Da$ = 5. (a) Case I, (b) II, (c) Case III, (d) Case IV.

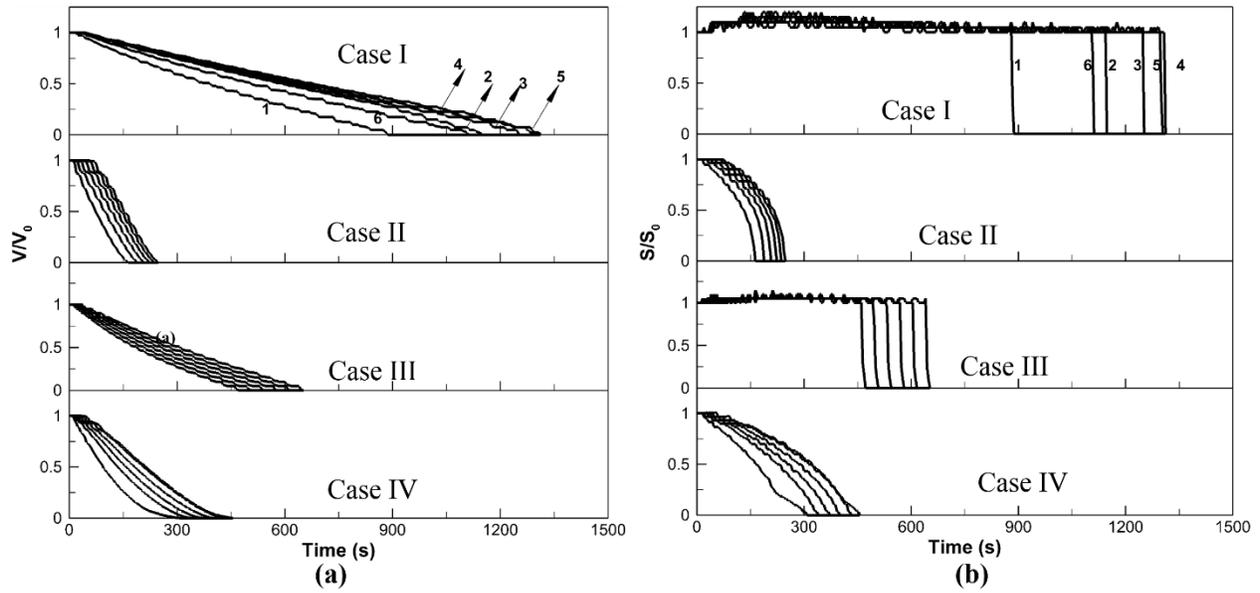

Fig. 7 Evolution of volume (a) and surface area (b) of mineral α in each elementary block with time during diffusion-controlled process with fast flow.

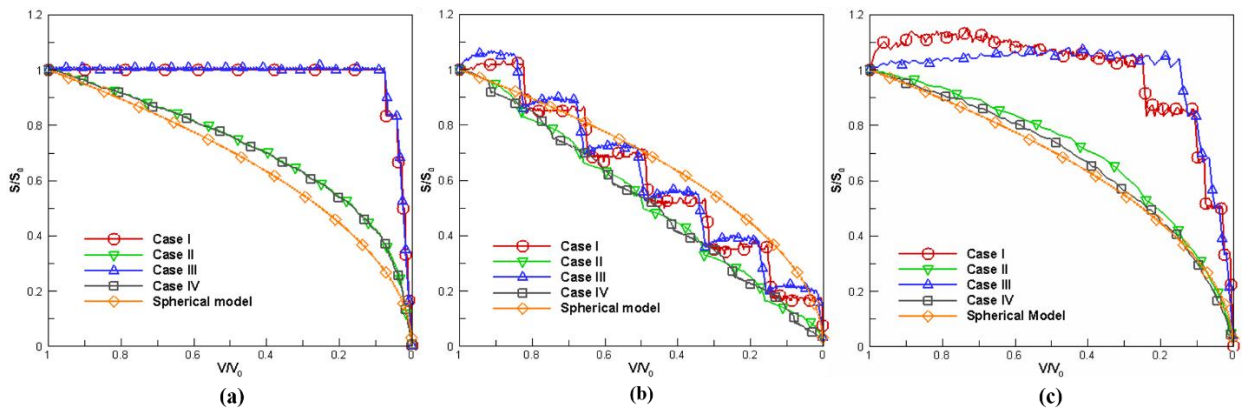

Fig. 8 Relationship of total reactive surface area and total volume of mineral α for different reactive transport process. (a) reaction-controlled process, (b) diffusion-controlled process with slow flow, and (c) diffusion-controlled process with fast flow.

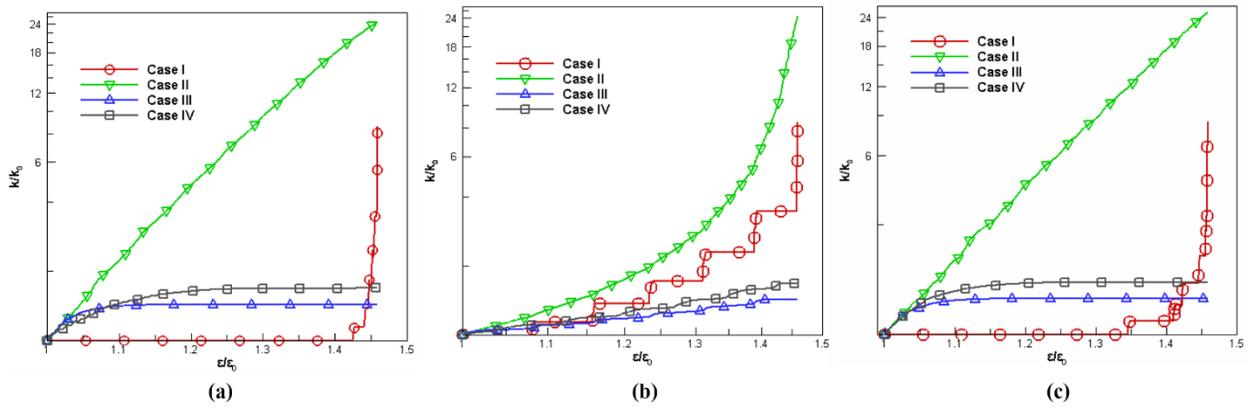

Fig. 9 Relationship of permeability and porosity during three typical dissolution process. (a) reaction-controlled process, (b) diffusion-controlled process with slow flow, (c) diffusion-controlled process with fast flow.

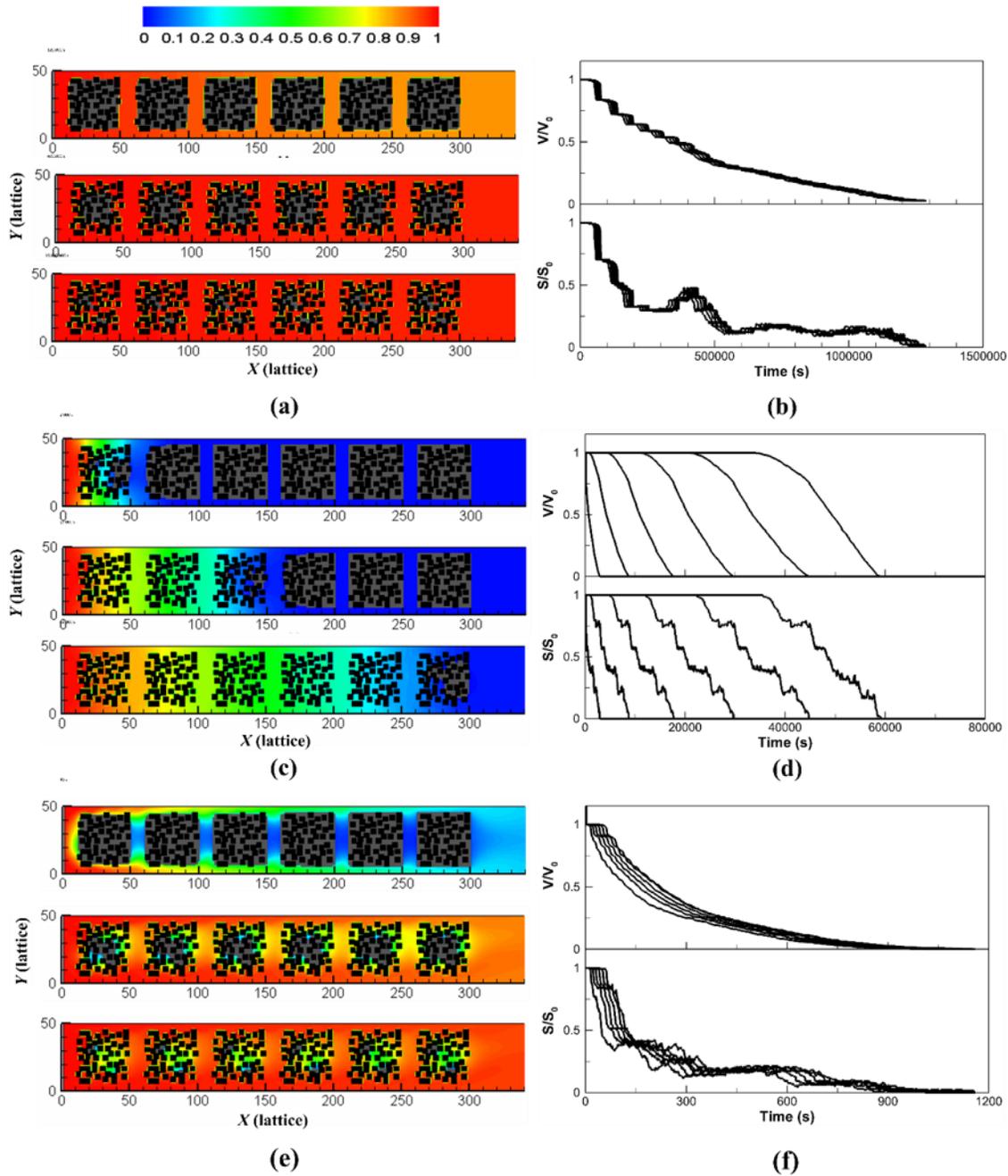

Fig. 10 The dissolution process of elementary blocks with random-distributed insoluble mineral B during three typical dissolution processes. (a, c, e) rectant concentration distribution and morpholoy of blocks at the beginning, middle and the last stages during reaction-controlled process, diffusion-controlled process with slow flow and diffusion-controlled process with fast flow, respectively. (b, d, f) evolution of volume of mineral A, reactive surface area and reactant concentration of each elementary blocks relative to the left figures (a, c, d).

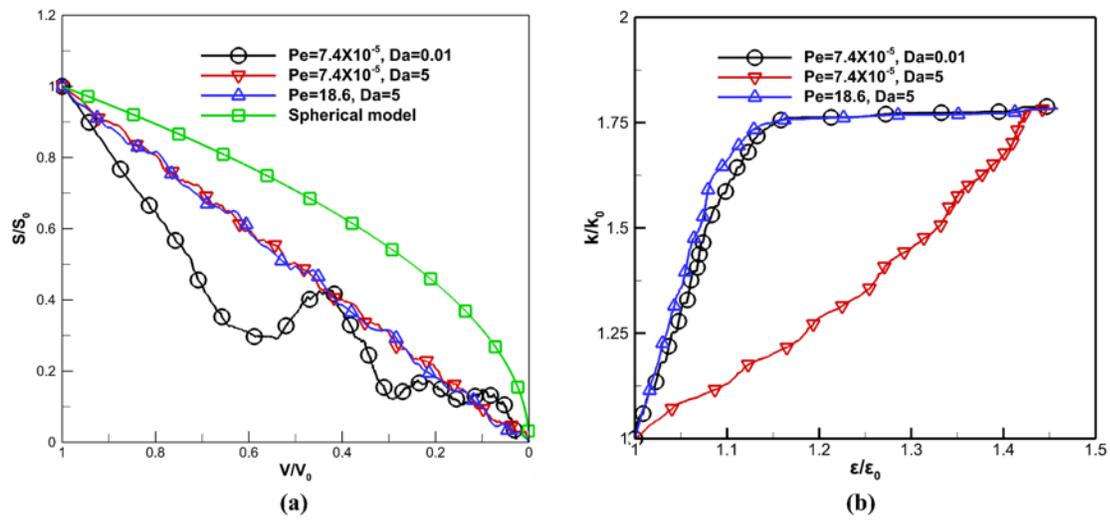

Fig. 11 Relationship of reactive surface area and volume (a) and relationship between permeability and porosity (b) during dissolution of elementary blocks with random-distributed insoluble mineral B.